# Quand rechercher c'est faire des vagues. Dans et à partir des images algorithmiques

**Pour citer**

Gaëtan Robillard, « Quand rechercher c'est faire des vagues : Dans et à partir des images algorithmiques » (prépublication), Recherche-création à la Cité Descartes (titre provisoire), Collection Across, Éditions Delatour, Sampzon, 2023.

## Résumé

*En recherchant la vague* est un film de synthèse réalisé en 2013, mettant en perspective le calcul de l'image par la simulation informatique, et par le texte et la voix. Prenant sa source dans une projection du film à l'Université Gustave Eiffel, l'article présente une réflexion sur la recherche-création dans et à partir des images algorithmiques. Au fond, qu'est-ce qui dans cette recherche-création – en particulier dans la recherche sur l'image algorithmique – peut être mis en mouvement ? Sans pour autant départager tout à fait ce qui serait recherche d'un côté et création de l'autre, nous nous attachons à caractériser des formes, des esthétiques ou des théories qui participent de déplacements possibles. Le relevé de ces possibilités est précisément l'enjeu du texte : des mathématiques à l'image et à la visualisation, de la naissance de l'esthétique générative à l'écriture de code informatique lié à des œuvres pionnières (*recoding*), ou encore du recensement de nouvelles esthétiques à de nouvelles formes de production critique.

Mots clefs : mathématiques, image algorithmique, visualisation, esthétique générative, recoding

## Biographie

Gaëtan Robillard est artiste, chercheur et professeur associé. Il a publié plusieurs travaux sur la matérialisation de l'image algorithmique, et a reçu le soutien d'instituts majeurs comme le ZKM en Allemagne, ou encore le consortium MediaFutures (Horizon 2020, UE). Sa thèse est une recherche-création en images numériques et sciences de l'art (Université Paris 8, 2022). Son travail est exposé dans des lieux comme le Palais de Tokyo (Paris), le Hessel Museum (New York) ou le Pearl Art Museum (Shanghai).

Rechercher en art peut s'apparenter à une quête d'exactitude. Dans un texte publié en 1928, Paul Klee écrivait que l'exactitude couplée à l'intuition engendre les meilleurs résultats[1]. Ici, la recherche-création est pensée comme un mouvement entre arts et sciences. Au fond, qu'est ce qui dans cette recherche fait mouvement ? *En recherchant la vague* est un film de synthèse réalisé en 2013 et qui met en perspective le calcul de l'image dans la simulation informatique, et par le texte et la voix. Prenant sa source dans une projection du film à l'Université Gustave Eiffel[2], l'article présente une réflexion sur la recherche-création dans et à partir des images algorithmiques.

Sans pour autant départager tout à fait ce qui serait recherche d'un côté et création de l'autre, nous nous attacherons à caractériser des formes, des esthétiques ou des théories qui participent de déplacements possibles dans la recherche-création. Le relevé de ces *possibilités* est précisément l'enjeu du texte : des mathématiques à la visualisation, de la naissance de l'esthétique générative à la réécriture de code informatique face à des œuvres pionnières (*recoding*), et au sein d'esthétiques algorithmiques contemporaines. Comme nous le verrons, ces déplacements sont liés à des formats ou des cadres normatifs dans lesquels la recherche-création se produit et où il s'agira d'identifier les leviers.

La discussion sera présentée en trois points. Premièrement, deux dispositifs seront discutés ; ils ont pour sujet la vague en tant que motif. Tout d'abord, *En recherchant la vague* dépeint une vague calculée qui fait image. La forme des mathématiques y est en jeu, en tant que recherche sur l'image de synthèse. Ensuite, *La vague dans la matrice* montre un champ de vagues généré par un algorithme issu des sciences du climat. Ces deux dispositifs sont présentés comme *œuvres de recherches* : ils naissent au sein de laboratoires de mathématiques puis paraissent dans l'espace de l'exposition.

Deuxièmement, nous présentons une recherche sur l'origine de l'esthétique algorithmique dans l'esthétique générative telle qu'elle fut pensée par le théoricien Max Bense dans les années 1960 à Stuttgart. La pensée cybernétique comme celle de l'entropie ou celle des processus stochastiques, tout comme la programmabilité de l'objet esthétique – entraîne alors des façons de penser l'œuvre en rupture avec les modèles de l'époque. Les travaux de l'artiste pionnier du Computer Art Frieder Nake sont discutés en tant qu'éléments fondateurs de cette esthétique. Le *recoding* (littéralement le recodage) est une méthode employée ici pour l'étude critique et la recherche-création.

Troisièmement, afin de mieux contextualiser l'esthétique algorithmique, nous présentons le recensement d'œuvres récentes issues du champ de la création visuelle contemporaine. Cet état de l'art fait apparaître des enjeux renouvelés de l'image algorithmique sous des traits identifiables: encodage, système, mathématiques, arbre de décision, *deep learning*, … Par des exemples, nous chercherons à désigner là où de nouveaux modes émergents, et déterminerons la façon dont ces modes font écho aux enjeux de la recherche-création telle qu'elle est pensée ici.

---

[1] Klee P., « Exact Experiments in the Realm of Art », Spiller J., *Paul Klee Notebooks: The Thinking Eye*, Vol. 1, Londres, Lund Humphries, 1961, p. 68-71.
[2] *Recherche-création à la Cité Descartes*, Université Gustave Eiffel, 2021.

## Vagues

*En recherchant la vague*

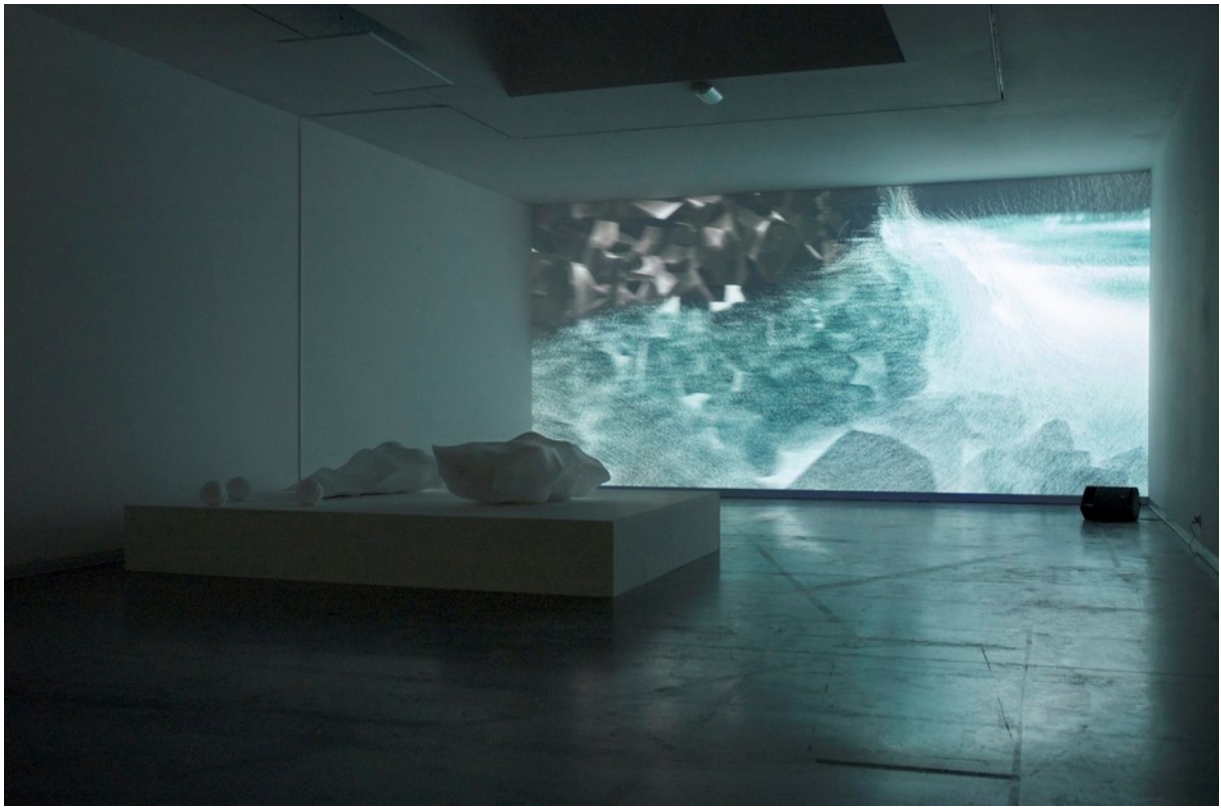

Figure 1 – *En recherchant la vague*, vidéo, lumière, sculptures en polystyrène fraisé, papier gaufré ; devant une simulation de fluides, les sculptures représentent les radeaux de fortune fabriqués par Henri Charrière pour son évasion ; vue d'exposition à l'Espace Culture, Université de Lille, 2013.

L'installation *En recherchant la vague*[3][4] (2013) s'inspire de l'autobiographie d'Henri Charrière[5]. Ce prisonnier du bagne de Cayenne, en Guyane française, utilisait des techniques élémentaires de comptage afin de se familiariser avec le mouvement des vagues pour ensuite organiser son évasion. L'installation comprend des sculptures et des images qui présentent la façon dont des équations mathématiques modélisent le mouvement et l'évolution de fluides dans le temps et dans l'espace.

Dans le film, alors que la caméra évolue lentement le long du rivage, le philosophe Bernard Stiegler demande (en voix off) : « Qu'est-ce qu'un domaine ? », puis soulève d'autres questions sur la profondeur, les constantes et inconnues issues des équations de Korteweg de Vries et de Navier-Stokes. Dans sa jeunesse, Stiegler a tenté de dérober une banque. Il a été arrêté et emprisonné pendant environ quatre ans. C'est dans ce contexte que son intérêt pour la philosophie a grandi. Il y a une superposition de cette figure hors champ qui lit le texte de la voix off, et la figure de Charrière qui hante l'image et qui néanmoins en est absente. Le texte de la voix off fonctionne comme sous-face de la computation, derrière l'image.

Précédant ou succédant au film, une lumière blanche révèle une série d'empreintes sur papier affichée au mur, révélant des traces de déchets trouvés sur le littoral, juxtaposées avec des traces d'équations. Lorsque le film s'achève, des sculptures apparaissent. Ces volumes ont été modélisés en 3D à partir du récit d'évasion d'Henri Charrière.

Du point de vue de l'image, la simulation informatique dépeint l'environnement de l'Île du Diable sur laquelle Charrière a été prisonnier. Une grille recouvre le fond marin. Les vagues composées de millions de particules entrent dans un mouvement qui semble infini. À travers la forme digitale de ces vagues – rendue particulièrement visible, des relations spatio-temporelles nouvelles se confondent avec les processus employés pour générer ces images. Pour mieux comprendre ces relations et ces processus, il est nécessaire de détailler la démarche de recherche et de création sur laquelle ils se fondent.

---

3 Crédits. Une production Le Fresnoy, Studio national d'art contemporain. Texte : d'après un entretien avec Caterina Calgaro et Emmanuel Creusé. Voix off : Bernard Stiegler. Image : Gaëtan Robillard, Vincent Audineau, Tabouret Studio. Son et création sonore : Rémi Mencucci, Paul de Robillard. Reportage vidéo : https://webtv.univ-lille.fr/video/6715/en-recherchant-la-vague-installation-de-gaetan-robillard
4 Robillard G., Calgaro C., Creusé E., « En recherchant la vague », *Rapport de recherche (Œuvres et recherche)*, Université de Lille, 2020.
5 Charrière H., *Papillon*, Paris, Éditions Robert Laffont, 1969.

Figure 2 – *En recherchant la vague*, recherches graphiques, équation de
Kazhikhov–Smagulov et objets trouvés sur des plages de l'Atlantique, 2013.

Dans ce travail, je me suis attaché à déconstruire le calcul et l'idée du calcul pour fabriquer l'image et le mouvement. C'est pourquoi il m'a paru nécessaire de me rapprocher de laboratoires de mathématiques, notamment dans le champ de la modélisation des phénomènes fluides. Impliqués dans la réalisation, les mathématiciens Caterina Calgaro et Emmanuel Creusé travaillent en effet sur la simulation numérique de tels phénomènes, comme l'évolution de gouttelettes d'eau dans l'air ou bien la propagation d'avalanches[6]. Les deux chercheurs étudient les équations correspondantes de Navier-Stokes et leurs variantes. Ils conçoivent alors des méthodes numériques de résolution, telles les méthodes de volumes et d'éléments finis, qui permettent de programmer l'ordinateur afin qu'il fournisse une représentation physique fiable des phénomènes.

L'un des processus important sur lequel se fonde ce travail est un dialogue au sein de notre équipe entre un artiste et deux chercheurs mathématiciens. Les rencontres qui ont rythmé ce dialogue ont permis d'opérer des transactions de sens entre des termes tels que « image », « modèle », « représentation », « réalité ». Le langage mathématique traduit en langage naturel est devenu une opération essentielle de la démarche artistique. Il en résulte un texte pour la voix off du film. Un travail graphique en gaufrage sur papier a également été réalisé à partir des équations liées à la modélisation des vagues et des flux, juxtaposées à des objets trouvés (Figure 2). Texte de la voix off et empreintes graphiques restituent la teneur conceptuelle et discursive du travail. L'idée d'image calculée se retrouve donc dans différents éléments de l'installation.

---

[6] Calgaro C., Creusé E., Goudon T., « Modeling and simulation of mixture flows: application to powder-snow avalanches », *Computers and Fluids*, n° 107, 2015, p.100-122.

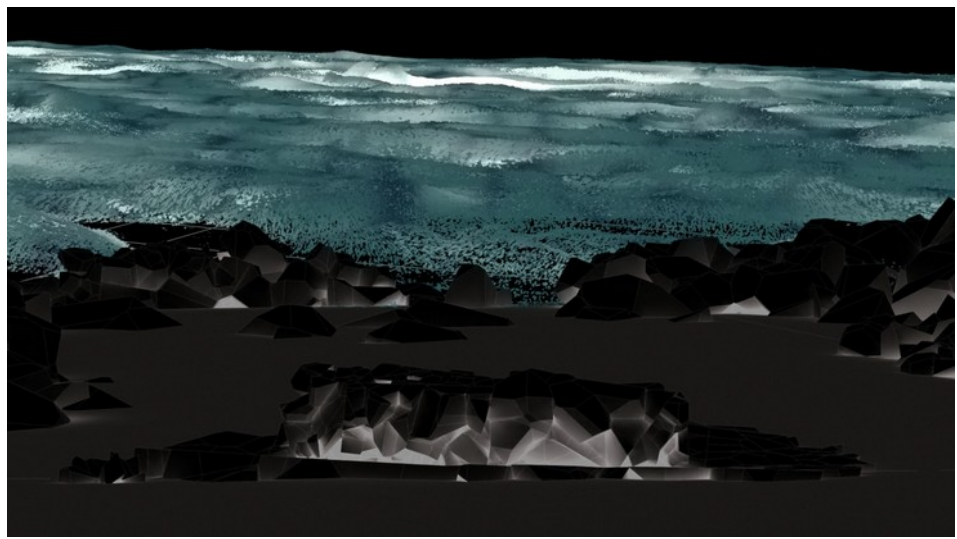

Figure 3 – *En recherchant la vague*, photogramme extrait du film, 2013.

Sous forme d'un plan séquence d'une durée de six minutes, le film – élément central du dispositif – résulte d'un travail de simulation de fluides[7]. Les images ont été calculées à partir d'un agencement spatial et mobile de particules dans une scène tridimensionnelle. Dans cet environnement, les particules sont animées grâce à des paramètres de forces physiques réalistes. Dans un domaine donné et d'une image à une autre, le logiciel calcule la position de chacune de ces particules selon que les forces varient entre elles (Figure 3). Les mouvements sont modélisés à partir de forces réalistes et paramétrables (gravité, friction, viscosité, vent, etc.). La simulation est tout d'abord pré calculée puis rendue en un film, image par image.

L'installation *En recherchant la vague* s'adapte selon le contexte d'exposition[8]. Le plan a souvent été modifié selon les contraintes architecturales, particulièrement la dimension de l'image projetée. La version idéale présente le film au fond d'une salle en longueur, couvrant toute la surface du mur du fond. Les sculptures sont présentées sur un grand socle (Figure 1).

Dans l'installation, la lumière est séquencée en précédant et en succédant le film de manière à prolonger le travail de l'image numérique au-delà de la surface de l'écran, dans l'espace vécu par le spectateur. Comme nous le verrons ensuite, le rapport entre image et calcul, entre une surface et une sous-face est un rapport qui s'inscrit dans un cheminement artistique lié à l'esthétique algorithmique, où le visible et le calculable sont co-opérants. Cette recherche nous a ensuite conduits à enquêter dans les sciences du climat, où la calculabilité et la visibilité des processus étudiés sont en jeu.

---

7 Avec le logiciel Houdini de SideFX.
8 Diffusions : Le Fresnoy (Tourcoing, 2013), Espace Culture Université de Lille 1 (2013), Palais de Tokyo (Paris, 2017), CCS Bard Museum, (New York, 2018), Akbank Sanat (Istanbul, 2019), Pearl Art Museum (Shanghai, 2019).

*La vague dans la matrice*

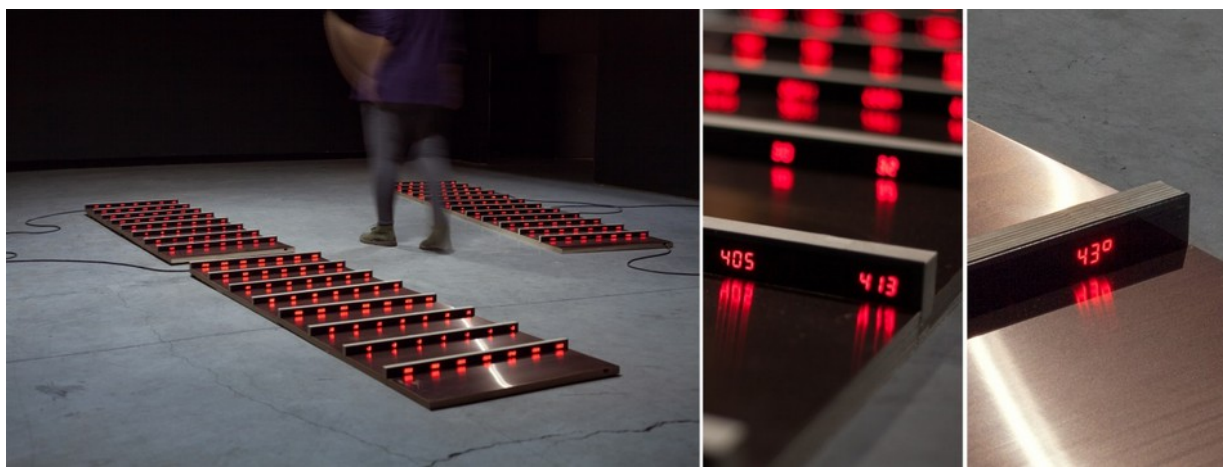

Figure 4 – *La vague dans la matrice*, 2019. À gauche : vue d'exposition, au centre : affichage des valeurs normalisées, à droite : affichage de la température de chacune des unités de calcul (32 unités au total).

*La vague dans la matrice*[9][10] (2019) a été développé à partir d'une immersion au sein du Laboratoire de Mathématiques et leurs Applications de Valenciennes (LAMAV) et du laboratoire de mathématiques Paul Painlevé. Ce travail examine la relation entre mathématiques et environnement, en particulier la façon dont les sciences du climat inscrivent de nouveaux signes dans le monde. Pour cela, l'installation se matérialise dans un réseau de surfaces en cuivre et de processeurs qui simulent un champ de vagues généré à partir de données océanographiques. La distribution du calcul rend visible un flux d'opérations qui à leur tour engendrent un déferlement de nombres. Toutes ces valeurs sont animées sur une grille d'afficheurs numériques disposée au sol.

Au cours du temps, les processeurs augmentent en température, et du fait de leurs limites matérielles, ils se désynchronisent progressivement. Le public assiste à une décomposition du mouvement du champ de vague, en fonction des échanges thermiques entre le matériel de l'installation et son environnement direct. L'affichage présente ensuite une série de températures qui tendent à décroître, tandis que le matériel se refroidit. Ce projet s'accompagne d'une performance électroacoustique[11] composée d'entretiens avec des scientifiques sur la submersion du littoral Atlantique en situation de tempêtes.

---

9 Crédits. Recherche en art : Gaëtan Robillard. Équipe scientifique : Caterina Calgaro (Laboratoire Paul Painlevé), Emmanuel Creusé (LAMAV), Bernard Fichaut (LETG), Jean-François Filipot, Rui Duarte, Andrea Ruju (France Energies Marine). Création sonore : Aude Rabillon. Conception Informatique et électronique : Martin Saez. Simulation stade maquette : Alexandre Rotolo. Soutiens : Résidence AIRLab ComUE Lille Nord de France, en partenariat avec le Fresnoy et Cinquante Degrés Nord, Festival RESSAC – UBO, avec le concours du CNRS. Documentation vidéo : https://vimeo.com/manage/videos/386770591
10 Robillard G., Calgaro C., Creusé E., « La vague dans la matrice », *Rapport de recherche (Œuvres et recherches)*, Université de Lille, 2020.

Le processus de création de *La vague dans la matrice* a débuté par une phase à caractère didactique. Nous nous sommes interrogés sur la signification physique des phénomènes dissipatifs (comme les écoulements de fluide) et leur traduction en langage mathématique. Les chercheurs mathématiciens ont transmis des savoirs généraux sur la mise au point d'un code de calcul numérique assurant la simulation d'écoulements de fluides. Une grande quantité de nombres et de multiplications matricielles sont nécessaires pour décrire la complexité du mouvement des vagues, des avalanches, de la pollution et des flux thermiques (Figure 6).

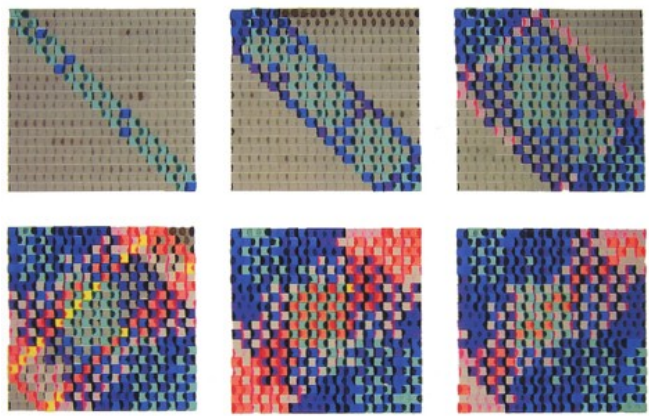

Figure 5 – Frieder Nake, *Matrizenmultiplikation, Serie 34* (détail), encre sur papier, généré par ordinateur, 50 × 50 cm, 1968.

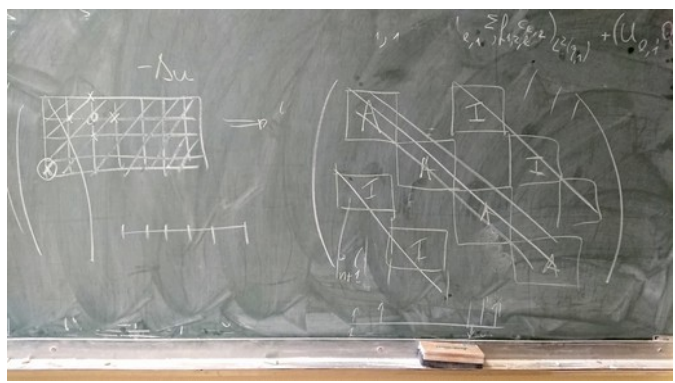

Figure 6 – Travail au tableau lors d'une phase d'immersion au Laboratoire Paul Painlevé, Université de Lille, 2019.

11 Robillard G., Rabillon A., *La base matérielle*, 2019. Réalisé dans le cadre des quatre-vingts ans du CNRS.

Nous avons ensuite travaillé sur des programmes qui font intervenir les équations de Navier-Stokes (en langage Freefem++) et de Korteweg de Vries (en langage Python). La simulation de tels modèles engendre à chaque itération le calcul de produits matrices-vecteurs. Le projet a alors entraîné une recherche sur la visualisation de telles matrices dans ce cas de simulation de vagues[12]. Par ailleurs, ceci a entraîné des échanges sur les méthodes numériques et les applications les plus récentes développées au sein du LAMAV et du laboratoire Paul Painlevé, par exemple pour la simulation d'écoulements à faible Mach[13]. Par ailleurs, la recherche sur la calculabilité des vagues s'est aussi étendue à une enquête au sein de laboratoires d'océanographie à Brest.

Au Laboratoire d'Océanographie Physique et Spatiale (LOPS) de l'Institut Universitaire Européen de la Mer, les états de mer sont modélisés de façon statistique. Notons que la physique statistique est très différente de la physique newtonienne sur laquelle les travaux du LAMAV sont fondés. Si travailler en immersion dans le LAMAV a permis une compréhension des enjeux computationnels dans le calcul de phénomènes physiques (où le rapport théorique entre cause et effet est premier), l'approche probabiliste du LOPS a ouvert notre recherche à une vision de la vague en tant que phénomène contingent. Nous nous sommes donc intéressés dans un second temps à un programme génératif de simulation d'un champ de vagues (en langage Python) à partir d'un spectre directionnel mesuré en condition réelle[14].

Le programme génératif a finalement été adapté à nos besoins et intégré à l'affichage des modules sculpturaux en cuivre. Les données considérées (hauteur significative, spectre directionnel) ont été enregistrées par la bouée CANDHIS (Les Pierres Noires, Atlantique Nord) sur le mois de février 2014, un mois marqué par des cyclones historiques liés au réchauffement climatique. Finalement, et à partir du traitement de ces données, l'installation génère un champ de vagues et distribue un calcul matriciel de hauteurs sur l'ensemble de trente-deux processeurs synchronisés[15], répartis sur les huit modules du dispositif. Les hauteurs obtenues en mètres sont ensuite normalisées afin de restituer de très grands nombres (Figure 4).

Car le calcul est distribué au sein de l'installation et que la fréquence de ce calcul sur chacun des processeurs est affectée par le seuil thermique du matériel lui-même, la temporalité des opérations diffère légèrement d'un processeur à l'autre. Dans ces conditions, l'unité spatio-temporelle de la matrice de vagues tend vers le désordre. Le système est entropique et devient progressivement asynchrone. Une fois la séquence de calcul terminée (après un nombre d'itérations déterminé), les objets de l'installation affichent chacun à leur tour une température. L'ensemble de ce processus (calcul du champ de vague en mouvement, arrêt et affichage des températures) se boucle sur une durée de vingt minutes environ. Chaque exposition est aussi l'occasion de développer une nouvelle configuration spatiale des modules qui constituent ce travail[16].

---

12 Robillard G., « La vague dans la matrice », *Journée Visu 2019*, Télécom ParisTech, 19 avril 2019.
13 Calgaro C., Colin C., Creusé E., « A combined Finite Volumes – Finite Elements method for a low-Mach model », *Int. J. for Numerical Methods in Fluids*, Vol. 90, n° 1, 2019, p. 1-21.
14 Ce programme émane d'un partenaire du LOPS, l'institut France Énergies Marines.
15 Raspberry Pi Zero W, 1GHz single-core CPU, 512MB RAM.

Plutôt que de restituer une image, l'installation présente un mouvement sous forme de valeurs numériques. Cette abstraction permet de mettre en avant le processus d'interprétation des données scientifiques en lui-même, tout en ramenant le spectateur à la situation dans laquelle l'opération se produit. Là où la mathématique devrait présenter une forme immatérielle, la température du *hardware* révèle l'inscription du calcul dans un espace physique.

En écho du minimalisme américain, le dispositif renvoie au cadre architectural, social et institutionnel dans lequel le dispositif s'inscrit. *La vague dans la matrice* est pensé comme *œuvre de recherche*[17], constituée à partir d'une enquête en laboratoire, pour interroger la nature de la calculabilité et le cadre interprétatif de données scientifiques. Enfin, signalons que sa forme et sa technique – leur généralisation en tant que machine – donneront lieu à un nouveau travail titré *Critical Climate Machine,* exposé au ZKM de Karlsruhe[18] et à l'Ircam Centre Pompidou.

Dans les années 1960, influencés par les théories de Max Bense, des mathématiciens et artistes comme Frieder Nake, Georg Nees et A. Michael Noll cherchèrent eux aussi à produire des résultats esthétiques à partir de méthodes de calcul informatique. C'est dans l'œuvre de Frieder Nake qu'on trouve en particulier des images abstraites qui traitent de la matrice en tant que structure mathématique et forme artistique (Figure 5). Comme nous le verrons dans la partie suivante, l'étude de cette iconographie révèle la dynamique de la relation entre image et algorithme, entre surface et sous-face[19].

## Esthétique générative

---

[16] Diffusions : Université Bretagne Ouest (Brest, 2019), Espace Croisé (Roubaix, 2019), Centre d'arts Ronzier de l'Université Polytechnique Hauts-de-France (Valenciennes, 2020).
[17] En référence au programme de résidence AIRLab (Université de Lille) et à la plateforme « Œuvres de recherches : Relais des collaborations Art/Sciences-Technologies en Hauts-de-France et Belgique » (Laboratoire Cristal, Université de Lille.
[18] Robillard G., « Critical Climate Machine », *Who Can we Trust?*, exposition collective, ZKM, 2022. Reportage vidéo : https://vimeo.com/667971904
[19] Nake F., « Think the Image, Don't Make It! On Algorithmic Thinking, Art Education, en Re-Coding », *J. of Science and Technology of the Arts*, Vol. 9, n° 3, 2017, p. 22- 31.

*Max Bense*

Max Bense (1910-1990) a étudié les mathématiques, la physique, la géologie et la philosophie à l'université de Bonn où il a obtenu son doctorat en philosophie et en science[20]. Philosophe, essayiste, poète, éditeur, collectionneur, il est aussi devenu une figure médiatique s'opposant à la société conservatrice dans l'Allemagne d'après-guerre. En 1960, alors qu'il est maître de conférences à l'université de Stuttgart, Max Bense publie *Aesthetica IV : Programmer le beau* [*Programmierung des Schönen*][21]. Ce quatrième livre achève un développement philosophique que Bense démarre en 1954, sur la voie d'une esthétique rationnelle. La théorie esthétique de Max Bense influence largement la scène pionnière du Computer Art des années soixante.

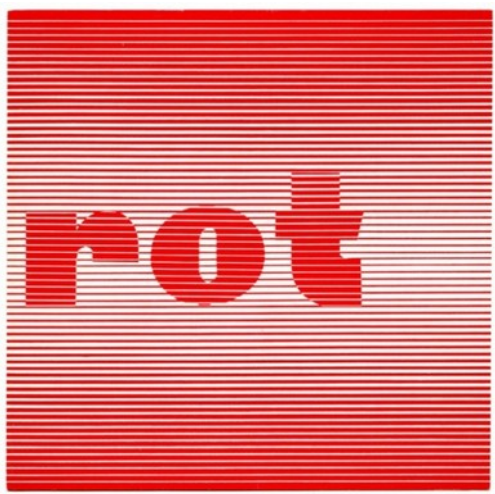

Figure 7 – Hansjörg Mayer, *Rot*, 1965, design de la couverture, ZKM | Center for Art and Media Karlsruhe / Elisabeth Walther-Bense Estate / ZKM-01-0129-02-0991-a.

Parallèlement à la publication d'*Aesthetica IV* en 1960, Max Bense et Elisabeth Walther-Bense lancent la revue *Rot* à Stuttgart. Elle réunira progressivement poésie concrète, sémiologie, typographie et cybernétique. Un *flyer* datant de 1966 présente la revue ainsi :

> La série Rot est destinée à la publication de littérature et de travaux graphiques expérimentaux. Le terme d'expérimentation n'est pas défini de façon restrictive. Nous incluons tout ce qui peut être créé à partir de l'hypothèse d'un concept esthétique théoriquement accessible, y compris le contenu de la forme et le but de la production artistique. La littérature et les travaux graphiques expérimentaux se rapportent à toutes les techniques topologiques abstraites et concrètes, stochastiques et aléatoires. De plus, [la série] ne se limite pas aux méthodes de production naturelles d'un individu créatif mais compte également avec les productions artificielles de systèmes informatiques électroniques.[22]

---

20 Walther E., « Max Bense's Informational and Semiotical Aesthetics », *Stuttgarter Schule*, 2000. [consulté le 2 novembre 2019]. https://www.stuttgarter-schule.de/bense.html
21 Bense Max, *Aesthetica IV : Programmierung des Schönen. Allgemeine Texttheorie und Textästhetik*, Baden-Baden, Agis, 1960.

Avec son format carré et sa couverture rouge conçue par Hansjörg Mayer (Figure 7), *Rot* est reconnaissable au premier regard. Marquée par un usage de la typographie Futura sans majuscules, la revue compte cinquante-sept cahiers publiés de 1960 à 1997. L'édition incarne les activités du Cercle d'intellectuels de Stuttgart qui s'est formé autour de Max Bense et Elisabeth Walther-Bense. La typologie de publication est riche tant pour les textes que pour les images : poésie en vers, prose, poésie concrète, philosophie, sémiologie, essai, théâtre, photographie, peinture, dessin, typographie, bande dessinée, diagramme ou texte généré par ordinateur...

La revue se consacre principalement à la poésie concrète et alterne fréquemment avec des numéros dédiés à la sémiologie et aux arts visuels. Trois cahiers, le n° 6, le n° 19 et le n° 50, sont explicitement orientés vers la programmation des formes esthétiques, avec respectivement des textes générés par Max Bense (1961), des dessins génératifs de Georg Nees accompagnés d'un texte de Bense prenant l'allure d'un manifeste (1965), et la poésie visuelle et programmée de l'Américaine Carole Spearin McCauley (1972). Il faut également noter que le n° 8 présente un texte d'Abraham A. Moles intitulé « Premier manifeste de l'art permutationnel », accompagné de la reproduction d'une image de Vasarely[23] (1962).

Sur le plan de l'image, ce qui transparaît à travers l'étude du matériel éditorial est avant tout une acceptation du monde visuel dans toutes ses nuances, allant de l'abstraction du signe typographique jusqu'à l'empreinte photographique du monde réel. Mais quel a été le rapport ou la posture du Bense-éditeur avec le matériel visuel produit *artificiellement* avec l'aide du système informatique ? Les archives du ZKM révèlent les maquettes de divers numéros, dont le numéro n° 19, rendu célèbre pour son inauguration de l'*esthétique générative* ; une première pour la publication des images algorithmiques dans le monde de l'art.

Publié en février 1965 sous le titre *Computer-Grafik*[24], le n° 19 se décompose en deux parties. La première se consacre à un texte et à un ensemble de dessins informatiques de Georg Nees, ingénieur-mathématicien et artiste travaillant alors chez Siemens à Erlangen en Allemagne. D'un point de vue historique, ces dessins comptent parmi les premières images générées par des algorithmes programmés sur un ordinateur[25]. En parallèle de cette publication, Bense organise un évènement public à la *Studiengalerie* de l'université de Stuttgart. Le philosophe y présente les dessins de Nees. Frieder Nake, également mathématicien et artiste pionnier du Computer Art, rapporte l'évènement comme étant la première exposition de Computer Art, « deux mois avant la célèbre exposition [de la galerie] Howard Wise à New York »[26].

---

22 Traduction de l'auteur. Allemagne, Karlsruhe, ZKM | Center for Art and Media, Elisabeth Walther-Bense Estate, ZKM-01-0129-02-0988, Bense M. et Walther-Bense E., feuillet publicitaire, 1966.
23 Moles A., *Erstes manifest der pemutationellen kunst*, Rot n° 8, Stuttgart, Max Bense et Elisabeth Walther, 1962.
24 Nees G., Bense M., *Computer-grafik*, Rot n° 19, Stuttgart, Max Bense et Elisabeth Walther, 1965.

Dans les deux premières pages de la revue, Georg Nees présente un texte « sur les programmes d'infographie stochastique ». Chaque graphique se compose à partir de paramètres aléatoires. La répétition de *paramètres aléatoires* « produit l'improbabilité esthétique des graphiques ». Après un court paragraphe, Nees présente cinq parties écrites sous la forme d'un pseudo-code. Chaque partie concerne une ou deux images que l'on découvrira dans les pages suivantes : « 8-sommet: (image 1) », « 23-sommet: (image 2) »[27], etc., jusqu'à l'image n° 6. En texte clair – sans spécification d'aucun langage de programmation – les descriptions restituent des instructions succinctes et programmables pour générer des résultats graphiques.

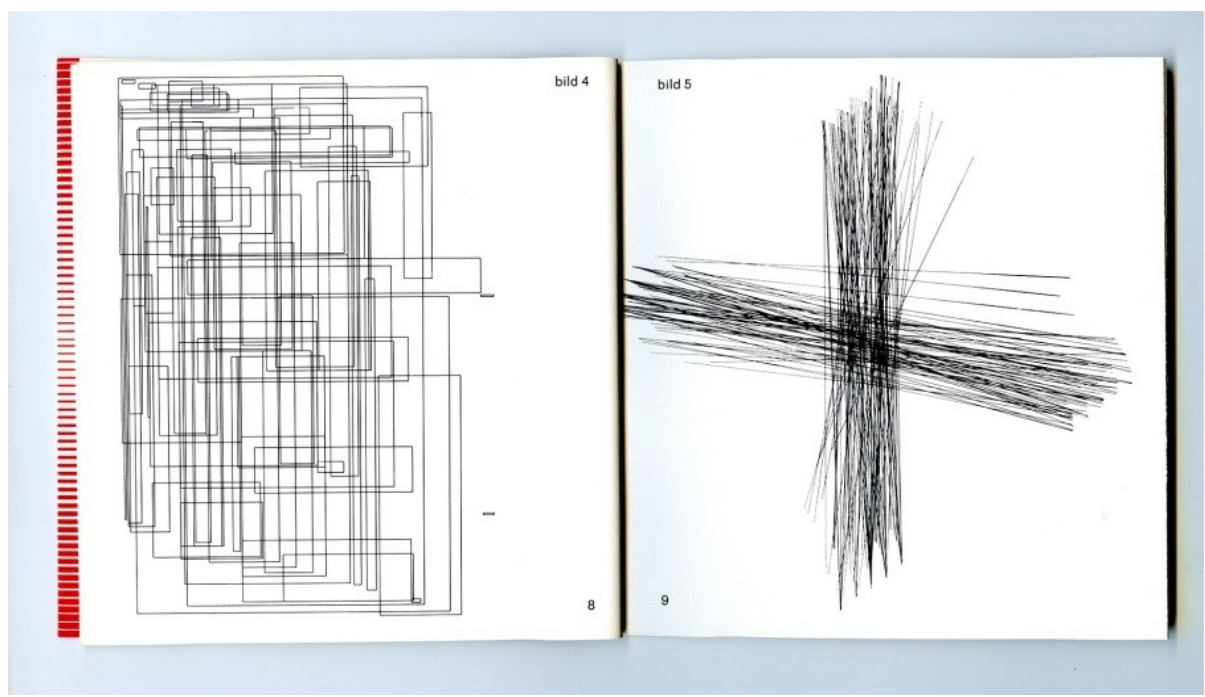

Figure 8 – Georg Nees, *Computer-Grafik*, in Rot n° 19, 1965, ZKM | Center for Art and Media Karlsruhe / Elisabeth Walther-Bense Estate / ZKM-01-0001-W-1014-f.

---

25 « Computer-Grafik (Nees 1965) », Nake F.(dir.), *Compart Center of excellence digital art*. [consulté le 17 novembre 2019]. http://dada.compart-bremen.de/item/exhibition/325
26 L'évènement public à la *Studiengalerie* et la publication de Rot n ° 19 datent de février 1965.
27 Traduction de l'auteur. Nees G., Bense M., *Computer-grafik*, *op. cit.*

Pour les images qui suivent, la mise en page est systématique : le titre de l'image (par exemple « bild 1») ; l'image elle-même, en pleine page et de haut en bas ; et la pagination. Les images présentées sont toutes faites de segments tracés à l'encre noire avec un trait d'une épaisseur d'environ 0,5 mm. Ce sont des lignes verticales, horizontales ou obliques. Les formes sont organisées sous forme de motifs dans une grille apparemment répétitive, ou bien en pleine page, à partir d'une variation de densité de lignes formant une composition abstraite (Figure 8).

La seconde partie du n° 19 fait paraître les « Projets d'esthétique générative » de Max Bense. Souvent présenté comme un manifeste des débuts du Computer Art, ce court texte de trois pages entend clarifier une méthode expérimentale de production d'états esthétiques par génération computationnelle[28]. Bense s'appuie sur la grammaire générative de Chomsky et se focalise sur la réalisation « artificielle » d'une structure esthétique programmable. Si à travers ses parutions, Bense conçoit la catégorie nouvelle des images algorithmiques, l'étude de la revue montre que sur ce terrain, l'activité du philosophe est indissociable d'une interaction au sein d'un groupe d'intellectuels et d'artistes : le Cercle de Stuttgart. Ses théories influencent de jeunes chercheurs comme Frieder Nake, qui au sein des laboratoires universitaires ou industriels, expérimentent et programment les premières images algorithmiques.

*Frieder Nake*

Frieder Nake est étudiant et chercheur en mathématiques au début des années soixante. Il assiste aux conférences de Bense et s'introduit dans le Cercle de Stuttgart. En novembre 1965, neuf mois après la publication et la présentation des expériences graphiques de Nees à la *Studiengalerie* de Bense, Nake et Nees exposent ensemble leurs travaux à la galerie du libraire Wendelin Niedlich [*Niedlichs Buchladen und Galerie*]. Cette exposition marque le début de carrière artistique de Nake, allant jusqu'à une reconnaissance internationale dès 1968[29].

L'œuvre de Nake entre 1963 et 1970 présente de nombreux dessins générés par des algorithmes, utilisant l'aléatoire comme modèle de l'intuition artistique. Pour comprendre les implications de l'esthétique générative de Bense et la façon dont un ensemble de productions visuelles lui font écho, nous avons décrit et analysé plusieurs images pionnières dans l'œuvre de Nake. Étant donné la nature computationnelle de cet objet d'étude, nous faisons appel à une approche empirique par le code. Ce principe expérimental et critique révèle d'une autre façon les processus à l'œuvre.

---

28 « Par esthétique générative, on entend la récapitulation de toutes les opérations, des règles, théorèmes, dont l'application à toute une série d'éléments matériels faisant office de signes rend possible par l'intermédiaire de ces derniers la production délibérée et méthodique d'états esthétique (qu'il s'agisse de répartitions ou de configurations) ». Bense M., *Aesthetica : Introduction à la nouvelle esthétique,* Paris, Les éditions du Cerf, 2007, p. 453.
29 Avec les expositions *Cybernetic Serendipity* à Londres, et *Tendencies 4: Computers and Visual Research* dans le cadre de la biennale *New Tendencies* à Zagreb. En 1970, le travail de Nake est présenté dans l'exposition collective de la 35ᵉ Biennale de Venize : *Ricerca e Progettazione : Proposte per una esposizione sperimentale* (Recherche et Design – Propositions pour une exposition expérimentale).

L'hypothèse est qu'un tel travail révèle l'unité de la pensée visuelle et de la pensée algorithmique dans ce qui fait image, particulièrement dans l'œuvre de Nake, mais aussi d'une façon générale dans l'image algorithmique. La pensée algorithmique entraîne une nouvelle façon de voir. L'étude combine donc une description visuelle et une description algorithmique, suivie d'une analyse générale, et de la partie expérimentale de l'étude : le *recoding*[30]. Nous nous intéressons ici en particulier au *recoding* comme activité de recherche.

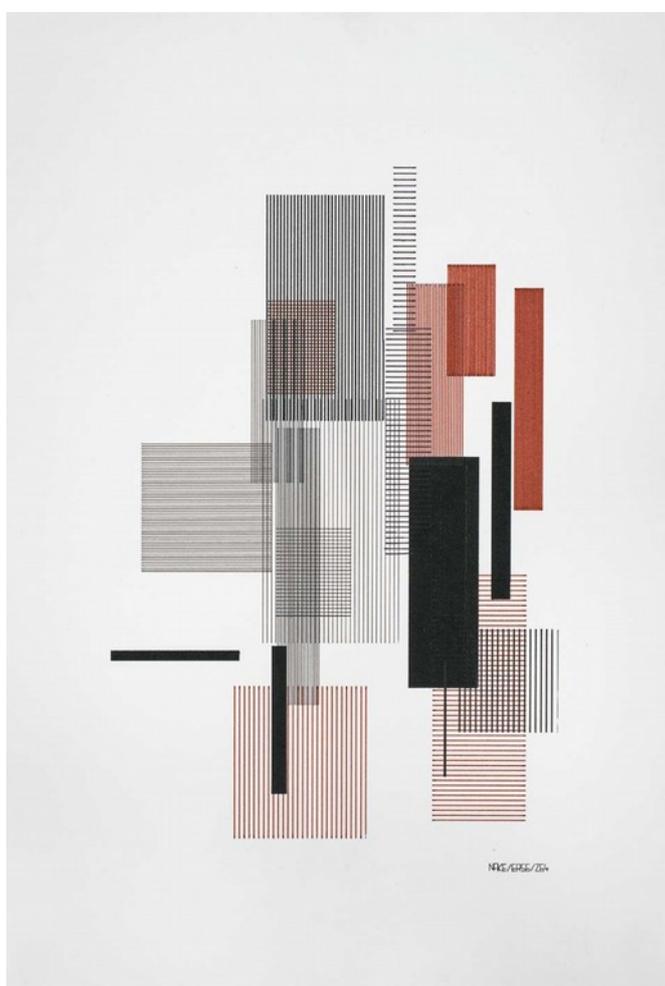

Figure 9 – Frieder Nake, *Rechteckschraffuren n° 3*, 30 mars 1965, infographie couleur, dessin au traceur et à l'encre de Chine (noir, marron) sur papier, 19.2 x 29 cm, Brême, Kunsthalle, Herbert W. Franke.

---

30 Que nous pourrions définir comme activité d'écriture d'un code informatique capable de reprendre des éléments algorithmiques ou visuels d'une œuvre algorithmique existante.

La conception d'algorithmes convoque une description formelle d'un processus. La description et l'analyse de l'objet d'étude sont soutenues par l'expérimentation d'un code exécutable, il en résulte donc une forme d'action. Autrement dit, l'approche discursive donne lieu à la programmation et à la réalisation de processus identifiés dans l'œuvre étudiée. S'appuyant à la fois sur des publications et de nombreux entretiens avec Nake, cette façon d'appareiller la recherche par le code vise à mieux comprendre la mathématique, les modèles ou les gestes développés par l'artiste dans son œuvre. Ce qui suit restitue des résultats obtenus à partir de l'étude de l'œuvre *Hachures rectangulaires [Rechteckschraffuren] n° 3* de 1965 (Figure 9).

*Hachures rectangulaires n° 3* est une œuvre graphique au format portrait. Le dessin a été programmé, exécuté, puis réalisé sur papier grâce à une machine à dessiner de type *pen plotter* (table traçante). L'œuvre se compose d'un ensemble de vingt zones de hachures rectangulaires en bichromie (noir et ocre), qui se chevauchent ou se juxtaposent. La composition géométrique présente des amas géométriques plus ou moins denses, et plus ou moins désordonnés. Une sensation de transparence se dégage du rapport entre des densités de hachures variables. Ces différentes densités engendrent des contrastes et des chevauchements.

Comment décrire l'algorithme ? Bien que certains enjeux de la programmation peuvent être perçus – comme la répétition et la variation, le programme ne peut être envisagé seulement par l'étude visuelle. Pour comprendre le programme original, il est nécessaire de s'appuyer sur un matériel publié par l'artiste lui-même. En 1966, Frieder Nake participe à une exposition et à une publication au Centre de calcul allemand de Darmstadt. Dans la publication, Nake décrit précisément les éléments qui composent le programme. Ce texte sera ensuite publié en anglais dans le catalogue de l'exposition *Cybernetic Serendipity* en 1968 à Londres. De plus, en 1974, dans son ouvrage rétrospectif *Ästhetik als Informationsverarbeitung*, Frieder Nake publiera les diagrammes et les algorithmes d'une grande partie de son œuvre.

À travers cette littérature, nous apprenons que le programme comprend un certain nombre de paramètres, ou variables aléatoires, qui correspondent à des caractères visuels. Par exemple, le nombre de hachures dans une zone de dessin est représenté par une variable déterminée. En outre, l'ensemble des valeurs des paramètres fixés au cours du calcul est obtenu à partir d'une distribution de probabilités appelée *F*. En termes mathématiques, une valeur aléatoire obtenue avec F est un évènement élémentaire parmi un ensemble d'évènements possibles. L'algorithme attribue une densité de probabilité pour chacun de ces évènements, et un générateur d'aléatoire est employé afin de sélectionner des nombres qui répondent à la densité voulue. Ainsi certains évènements ont plus de chance de se produire que d'autres. Autrement dit, bien que le résultat visuel semble chaotique, le désordre résulte en réalité d'un paramétrage précis. Le *recoding* permet de saisir les enjeux esthétiques de cette approche (Figure 10).

L'écriture et l'exécution de code propre à cette méthode fonctionnent comme une tentative de description des étapes de calcul ainsi que des différents niveaux d'abstraction de l'œuvre. On va de la la distribution de probabilités jusqu'à la réalisation du dessin. Le premier recodage, « Hachures », est un recodage élémentaire, où la loi de densité (ou la distribution) est uniforme. Dans ce recodage, les valeurs pour chacun des paramètres visuels (tels que position, largeur, hauteur, densité et direction des hachures) sont équiprobables. Si l'on superpose les résultats visuels issus de plusieurs exécutions du programme sur un même plan, la surface de travail tend à noircir entièrement.

Le second recodage, « distribution de densité » est un algorithme qui entraîne la distribution de probabilités à partir d'une approche discrète. Ici, l'algorithme est utilisé pour générer une grille de rectangles dont les valeurs de gris répondent à la distribution voulue. Si la distribution était uniforme, la surface visuelle présenterait des contrastes homogènes (aucune valeur de gris ne dominerait par rapport à une autre). Ici, le paramétrage est non uniforme. La distribution de densités entraîne des dominantes de valeurs. Dans la distribution voulue, les valeurs de gris moyens dominent tandis que les valeurs hautes et basses sont mineures. En somme, l'image est pensée comme une probabilité d'évènements dont chaque actualisation réalise un état possible compris au sein d'un très grand ensemble d'états. Cette expérimentation montre comment l'approche probabiliste de l'image engage un geste à la fois déterminant et ouvert. L'image algorithmique est une image conceptuelle et expérimentale par essence.

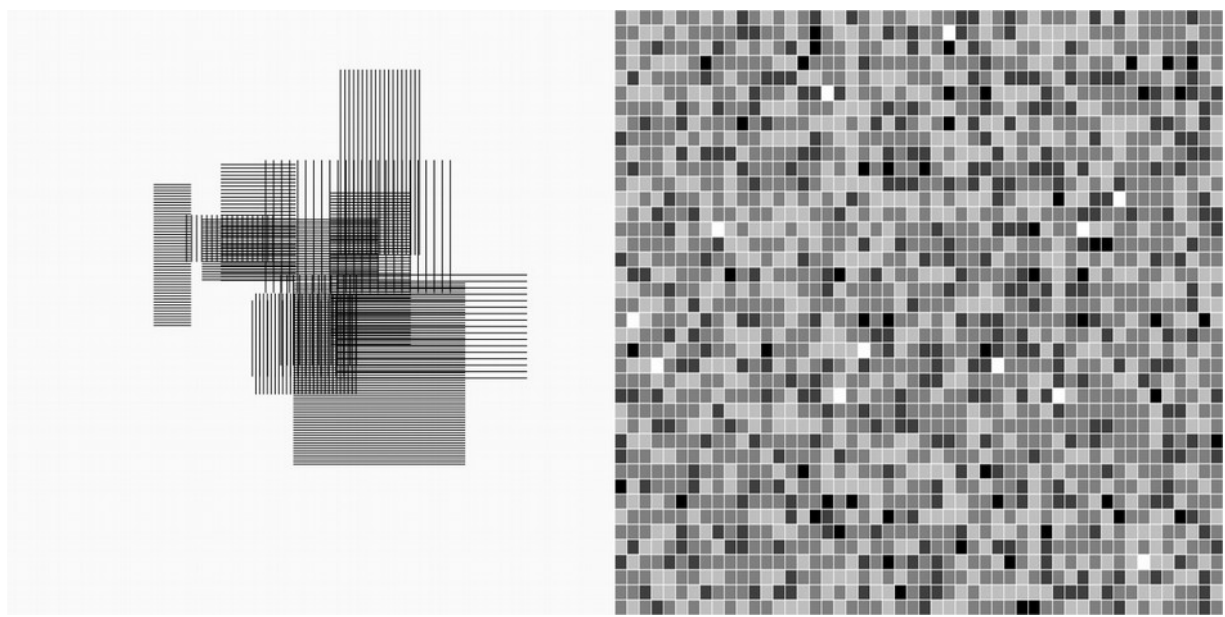

Figure 10 – « Hachures » et « distribution de densité », deux *recodings* en langage JavaScript (p5.js) à partir de l'œuvre Hachures rectangulaires de Frieder Nake.

Bien que programmée, la dynamique de l'image est bifurquée au sein même du programme par une combinaison entre variables aléatoires et définition de lois de probabilité. Chaque exécution du programme résulte potentiellement en un nouvel équilibre. Il faut aussi ajouter que l'algorithme, une fois programmé, est toujours paramétrable et reprogrammable *ad infinitum.* C'est aussi ce que le *recoding* permet de vérifier par itération et enrichissement successifs de l'algorithme. Comme le souligne Nake, l'œuvre est le résultat d'une sélection dans un espace potentiellement infini (il ajoutera dans ses travaux théoriques qu'en ce sens l'œuvre est « non-finie ») :

> La première tâche qui est aussi la plus importante est de préparer un programme qui doit permettre la production d'une classe entière de dessins (« des objets esthétiques », tels que mentionnés par Max Bense) parcourant un motif spécifique dans toutes ses déclinaisons. Une analogie peut être ici construite avec le processus artistique de la poursuite d'un thème à travers toutes ses possibilités, possibilité guidée par l'intuition. Ici, le concept d'intuition fait référence au choix des possibilités à partir d'un répertoire donné. L'ordinateur simule l'intuition par la sélection automatique de nombres pseudo-aléatoires.[31]

Avec l'étude de l'esthétique générative des années soixante apparaissent des relations fondatrices entre algorithme et image, entre information et originalité. L'intégration du hasard comme composant essentiel de l'esthétique générative implique une œuvre contingente. Influencée par Bense, l'œuvre de Frieder Nake est une œuvre ouverte qui allie pensée algorithmique et abstraction géométrique. Cette œuvre fondatrice pour les arts numériques est aussi une œuvre de recherche en cela qu'elle intègre la science des probabilités au service d'une nouvelle épistémologie algorithmique et visuelle.

---

31 Nake F., « Notes on the Programming of Computer Graphics », In *Cybernetic Serendipity: The Computer and the Arts*, Reichardt J. (dir.), Londres, Studio International, 1968.

Désormais, Nake théorise une œuvre-processus[32], déterminée essentiellement par la dynamique du changement. Ainsi la dialectique de l'image algorithmique paraît anticiper la nature complexe des images du XXIe siècle, en particulier dans le domaine de la visualisation et des médias de synthèse (*synthetic media*), notamment marqués par les algorithmes de l'apprentissage automatique. Dès lors, de quelle façon l'esthétique générative agit-elle dans le présent ? Dans la partie suivante, cette question nous amène à étudier la création artistique contemporaine, notamment du point de vue de l'articulation esthétique entre le visible et le calculable.

## Nouvelle esthétique ?

Nous entendons résumer ici un état de l'art de la création contemporaine (en art, design, architecture, cinéma, littérature...) où *la programmation d'algorithmes et le code* constituent une partie prépondérante de l'œuvre[33]. En particulier, nous centrons cette approche sur des œuvres qui sont faites d'images ou bien qui relèvent d'un régime génératif, et où des caractères informatiques ou computationnels sont identifiables. Sur le plan du cadre temporel, le recensement s'étend entre deux expositions notables du ZKM, Centre d'art et des médias à Karlsruhe : *Algorithmic Revolution* (2004), et *Open Codes* (2019).

Du discours de la rupture à celui de l'ouverture, comment évolue l'esthétique algorithmique ? Comment dans le début du XXIe siècle les artistes travaillent-ils avec les algorithmes ? Le cadre et les méthodes qui ont conduit le recueil des données sont présentés, puis trois œuvres extraites du recensement sont décrites.

L'état de l'art prend la forme d'un inventaire d'images et de textes brefs en français et anglais qui décrivent soixante-deux œuvres avec leurs algorithmes. L'inventaire est chapitré en quatre grandes parties : *dessin et code*, *image et temps*, *dit et écrit*, *matériel et externalité*. Ces quatre couples montrent l'étendue et la présence accrue des esthétiques algorithmiques dans les catégories ou formats traditionnels de l'art. Chaque partie fait l'objet d'introductions qui ne sont pas entièrement restituées ici. Grâce à cette structure, nous espérons pouvoir décrire différents régimes esthétiques et leurs évolutions.

---

32 Nake F., « Georg Nees & Harold Cohen: Re:tracing the origins of digital media », *op. cit*.
33 Cf. Robillard G., *Des algorithmes à l'œuvre : Naissance du Computer Art et environnements génératifs*, Thèse de doctorat en esthétique, sciences et technologies des arts, sous la direction de Alain Lioret et Thierry Mouillé, Université Paris 8, 2022.

En terme méthodologique, la rédaction et l'iconographie s'accompagnent du développement d'un outil éditorial léger écrit en code Pyhton et dont le but est la production d'un document de type Markdown (un langage de balisage léger pour créer du texte formaté (html, epub, docx, pdf, etc.) tout en utilisant un éditeur de texte simple). Exploitant les données recueillies, classées et annotées, cet outil a été pensé de façon à assister la mise en page à destination de différents formats de publication, papier et numérique.

D'autre part, une étude sur la description de la partie algorithmique des œuvres a été conduite. Dans un premier temps, un « algo-type » a été défini pour chaque œuvre. Étant donné leur caractère technique et non visible, les algorithmes des œuvres contemporaines sont rarement présentés par les artistes ou les lieux d'exposition. Pour les décrire précisément, il serait nécessaire de lire le programme informatique lui-même, c'est-à-dire le code source. Si dans nombre de cas il a été possible d'identifier le nom ou nature des algorithmes employés, il reste qu'une large partie de ces algorithmes ne peut être ni relevée, ni donc être catégorisée selon une taxonomie stricte.

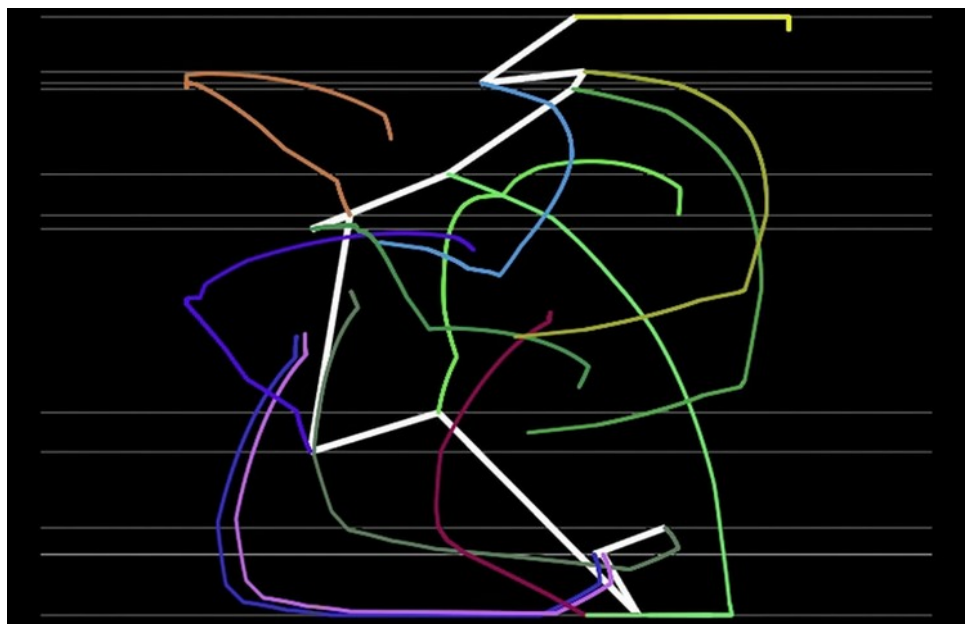

Figure 11 – Manfred Mohr, *P2200 (Traces)*, 2014.

C'est pourquoi il a été nécessaire – à travers les données recueillies – d'identifier des traits (ou attributs) qui aident à la description des œuvres du point de vue des algorithmes. Ces attributs caractérisent des opérations conduites au sein de la création de ces œuvres. Huit attributs ont donc été déterminés :

- *Encodage* ; lorsqu'il est évident que l'œuvre encode ou décode des données.

- *Système* ; ou appareillage, combinaison d'éléments matériels ou logiciel faisant système.
- *Mathématiques* ; lorsque la dimension mathématique est une partie essentielle de l'œuvre.
- *Arbre* ; lorsque l'algorithme fait apparaître un arbre de décision.
- *Deep Learning* ; ou apprentissage profond – correspondant à une sous-catégorie des algorithmes de l'apprentissage automatique (ou intelligence artificielle), lorsque l'œuvre emploie de tels algorithmes.
- *Interactivité* ; lorsque l'œuvre engage une interactivité avec le public.
- *Internet* ; lorsque l'œuvre repose sur ou exploite des données issues du web.
- *Plateforme* ; lorsque l'œuvre se confond dans un réseau de ressources diverses et qu'elle est orientée pour des usages en ligne.

Ces traits font suite à une réflexion sur la façon dont il serait possible de classifier le monde des algorithmes. Pour aider à la navigation dans les données selon les traits présentés, des pictogrammes ou glyphes ont été choisis. Chacun de ces glyphes souligne des attributs liés à l'œuvre. Nous espérons que ceci permettra d'effectuer ultérieurement de nouveaux tris dans les données et donc de favoriser la partie critique et épistémologique de la recherche. Notons que d'autres catégories ou attributs ont pu être envisagés sans que ceux-ci soient suffisamment récurrents dans l'ensemble pour être retenus : interface, visualisation, expérimental, archive, … Nous restituons ici la description de trois œuvres choisies de l'état de l'art.

[Encodage] *Sketch_150709b,* Mattis Kuhn, 2015

Comment mettre en relation code informatique et *output* ? Quelle est la pertinence de la méthode de fabrication lorsqu'elle n'est pas notée dans le produit final ? Quelle est la valeur du non-visible ? Pour tenter de répondre à ces questions, ce travail vidéo d'environ 12 minutes montre une séquence de captures d'écran. Elles présentent divers algorithmes écrits en langage de programmation Processing pour la créer chaque fois une image identique : un carré noir sur fond gris. Le premier algorithme se compose de trois lignes de code et montre la manière la plus simple et la plus courante de procéder. Des algorithmes plus complexes, particuliers ou abscons, des fonctions intégrées, des points, des lignes et des formes se succèdent pour aboutir strictement au même rendu visuel.

[Mathématiques] *P2200 (Traces)*, Manfred Mohr, 2014

Voici une série tirée du cycle de travaux Artificiata II (Figure 11). Cette série commence par un livre montrant les soixante-six projections 2D d'un chemin diagonal à partir d'un hypercube en douze dimensions. Ici, un chemin diagonal est une ligne à segments multiples traversant un hypercube à n dimensions, où chaque changement de direction indique le passage à travers une de ces dimensions. Dans *P2200* (habituellement présenté sur écran), l'épaisse ligne blanche montre la rotation d'un tel chemin, dénoté par des lignes horizontales, et projetées sur une surface. Les lignes courbes en couleur retracent ce mouvement. Décrites par Manfred Mohr comme des « êtres graphiques », les images passent d'une couleur à l'autre, créant ainsi différentes expériences visuelles qui pourraient être comparées, dans le domaine musical, à la polyphonie et à l'unisson.

[Encodage][Système] *Random Access Memory,* Ralf Baecker, 2016

Le dispositif présente une mémoire numérique entièrement fonctionnelle. Au lieu de fonctionner sur des composants semi-conducteurs pour représenter les états binaires de 0 ou 1, la mémoire utilise des grains de sable comme matériau de stockage. Les grains de sable peuvent être lus, placés ou retirés sur un disque rotatif, tandis qu'une caméra microscopique suit la position d'un seul grain. Ce mécanisme est sujet à des erreurs dues à des lectures erronées, à des grains qui rebondissent et à d'autres accidents. L'algorithme exécuté est un Turmite, une machine de Turing fonctionnant en deux dimensions. Son seul but est d'écrire autant de 1 (les grains physiques) que possible dans la mémoire et d'éviter une interruption.

Ainsi présentés, l'état de l'art et son étude ont permis de révéler de nouveaux aspects du rapport entre image et algorithme. Par exemple, le geste humain est confronté au geste de la machine (*Artist and Machine*, Susie Fu, 2018) ; l'archivage et la reproduction d'œuvres historiques du Computer Art est pensé par le biais de la retranscription du code (*The ReCode Project*. Matthew Epler, 2012) ; par le truchement du code, le dessin est pensé comme système en évolution où un état donné conditionne l'état suivant (*Life and Death of an Algorithm*, Troika, 2017) ; ou encore, l'algorithme fait apparaître en lui-même la dimension temporelle ou spatiale de la ligne – comme nous venons de le voir avec Manfred Mohr.

Qu'est-ce que tout cela implique ? Que peut-on dire de la contemporanéité de ce rapport, entre ce qui se voit et ce qui se calcule ? La ligne est un élément qui appartient au langage des précurseurs comme Frieder Nake ou encore Vera Molnar. Elle continue à être un sujet dans l'art algorithmique d'aujourd'hui. À contrario, l'aléatoire, qui caractérise tant les premiers pas du Computer Art joue un rôle moins déterminant. D'autres préoccupations sont apparues, comme celles de la description, de la transcription, ou de la construction des réalités contemporaines. Ceci est particulièrement vrai dans l'œuvre cartographique *Anatomy of an AI system* de Kate Crawford et Vladan Joler (2018), une description graphique du système Amazon Echo, ou encore avec *Algoffshore* du collectif RYBN (2018).

On peut observer que l'autonomie de l'abstraction géométrique et du formalisme algorithmique laisse place à d'autres stratégies où non seulement les algorithmes sont amenés à modéliser des processus de création – comme pour la recherche du dessin d'un nouveau personnage manga pour l'auteur défunt Osamu Tezuka, mais en plus engendrent des espaces algorithmiques ou des traces visuelles et conceptuelles à travers lesquels il est question d'installer voir de partager les pratiques elles-mêmes. Dans *Landscape Prediction: An Earthology of Moving Landforms (2018),* l'artiste Abelardo Gil-Fournier propose ainsi un atelier pour générer des vidéos prédictives de vues satellites des méandres de l'Amazone. Selon ses propres mots, le dispositif fonctionne comme

> une plateforme pour spéculer sur cet enchevêtrement particulier entre les médias visuels et les surfaces de la planète, au-delà des contextes extractifs et financiers qui l'ont fait naître.

Nous comprenons que le travail de l'artiste porte ici davantage sur le cadre médiatique de l'image algorithmique ou son inscription économique et environnementale que sur l'esthétique de l'objet comme on peut la trouver dans l'esthétique générative pionnière.

Pour citer David Zerbib à propos du format comme principe d'une esthétique contemporaine, « c'est bien dans le cadre matriciel de l'œuvre, dans le réglage de ce qui peut la produire, et dans ses conditions opératoires, que réside l'enjeu esthétique. » L'esthétique algorithmique contemporaine, qui s'éloigne du formalisme des pionniers, montre comment le code informatique, sa performativité, et les cadres qu'il implique deviennent les lieux même de nouvelles délinéations de l'œuvre d'art. Tout en souscrivant au format comme principe d'une esthétique contemporaine, nous soutenons que l'exploration des normes d'inscription de la recherche-création est un travail nécessaire pour mieux penser la recherche-création en tant que telle.

## Conclusion

Quels sont les leviers de la recherche-création telle qu'elle a été discutée ici ? Pour résumer, quatre axes ont été identifiés. Ils sont présentés ici:

- Le statut d'œuvre de recherche : une œuvre qui émane d'un travail entre art et science, en immersion dans des laboratoires, et dont la capacité est de déplacer l'objet de la recherche dans des *situations* singulières, c'est-à-dire hors des cadres scientifiques pré-établis. Dans le numérique, l'œuvre de recherche peut être généralisée en tant que machine ou *software* – qu'il soit écrit complètement, partiellement, ou bien qu'il soit *ready-made.*

- L'épistémologie algorithmique et visuelle fondée sur une approche probabiliste : l'étude des processus stochastiques au cœur du Computer Art pionnier permet un raisonnement critique sur l'image algorithmique ; le *recoding* soutient une telle épistémologie, car de façon empirique ce travail étend les possibilités d'analyse des œuvres algorithmiques autant que la compréhension de leur nature opérationnelle.

- La nécessité de penser des formats et des outils propres à la recherche-création, spécifiquement la recherche-création numérique : il apparaît par exemple que la publication d'un code source ou bien d'autres systèmes de notation permettrait une meilleure étude critique des objets de recherche, donc à de nouveaux formats de publication ou d'archivage.

- En somme, de nouveaux cadres d'inscriptions pour les objets de la recherche-création mériteraient d'être recherchés.

Ces axes, nous l'espérons, doivent permettre de futures discussions à un niveau interdisciplinaire. Considérons également que la recherche-création doit jouer un rôle pour affirmer le caractère essentiel de l'art, du design et des humanités dans la recherche – en particulier dans un monde sujet à l'automatisation accrue du travail intellectuel.